\newif\ifreview
\reviewfalse  

\documentclass[10pt,conference]{IEEEtran}
\usepackage{cite}
\usepackage{url}
\usepackage{adjustbox}
\usepackage[dvipsnames]{xcolor}
\DeclareRobustCommand{\visionrev}[1]{{\color{black}#1}}
\usepackage{graphicx}
\usepackage{balance}
\usepackage{booktabs}
\usepackage{amsmath}

\usepackage{mdframed}
\usepackage{etoolbox}
\makeatletter
\newcommand{\honestyrefcolor}[1]{\ifstrequal{#1}{martinez2022survey}{\color{black}}{\ifstrequal{#1}{verdecchia2023green}{\color{black}}{\color{black}}}}
\pretocmd{\@bibitem}{\honestyrefcolor{#1}}{}{}
\pretocmd{\@lbibitem}{\honestyrefcolor{#2}}{}{}
\makeatother
\usepackage{needspace}

\newcommand{\oururl}[1]{\textcolor{Black}{\texttt{\url{#1}}}}

\definecolor{thesisline}{HTML}{4B0082}
\newmdenv[
    nobreak=true,
    leftline=true,
    rightline=false,
    topline=false,
    bottomline=false,
    linecolor=thesisline,
    linewidth=1.5pt,
    leftmargin=0pt,
    rightmargin=0pt,
    innerleftmargin=8pt,
    innerrightmargin=0pt,
    innertopmargin=3pt,
    innerbottommargin=3pt,
    skipabove=5pt,
    skipbelow=5pt
]{thesis}

\begin{document}
%
\title{\visionrev{Honesty Testing of LLMs and AI Agents: A Vision for Software Engineering}
}

\ifreview

\newcommand{\repourl}{\texttt{\oururl{https://anonymous.4open.science/r/se4aisci}}}

\author{
\IEEEauthorblockN{Anonymous Authors}
\IEEEauthorblockA{
\\
}
}

\else


\author{
\IEEEauthorblockN{Ying Xiao\IEEEauthorrefmark{1}, Xinjian Wu\IEEEauthorrefmark{1}, Lukas Twist\IEEEauthorrefmark{1}, Yaoqi Guo\IEEEauthorrefmark{2},\\
Yepang Liu\IEEEauthorrefmark{3}, Zhenpeng Chen\IEEEauthorrefmark{4}, Jie M. Zhang\IEEEauthorrefmark{1}}
\IEEEauthorblockA{\IEEEauthorrefmark{1}King's College London, UK\quad \IEEEauthorrefmark{2}Nanyang Technological University, Singapore\\
\IEEEauthorrefmark{3}Southern University of Science and Technology, China\quad \IEEEauthorrefmark{4}Tsinghua University, China}
}

\fi


%


\maketitle


\begin{abstract}


\visionrev{Human oversight of large language models (LLMs) and artificial intelligence (AI) agents increasingly depends on their reports. Claims that conflict with a system's own judgments can conceal failures, prevent timely intervention, and contribute to severe harm. Greater autonomy makes faithful reporting increasingly important for maintaining human control. In this vision paper, we propose \emph{honesty testing of LLMs and AI agents} as a new research area in software engineering. Building on existing honesty evaluations, our goal is to make faithful reporting a software quality requirement that developers can specify, test, and maintain. We define \emph{honesty bugs} as behavioural defects in which an LLM or AI agent, when required to communicate faithfully, makes an explicit claim that conflicts with its own judgment at the time of reporting. We present an initial framework covering LLM-level testing under controlled inputs and agent-level testing of complete systems. The framework connects these levels to testing workflows and oracle evidence. We identify research directions in benchmarks, model and harness analysis, oracle construction, and lifecycle testing to support effective oversight alongside task correctness.}

\end{abstract}


\section{Introduction}

\visionrev{Following advances in large language models (LLMs)~\cite{brown2020language,ouyang2022training}, artificial intelligence (AI) systems increasingly operate as agents that reason, use tools, and process files~\cite{wang2024survey,plaat2025agentic}. Applications include scientific research~\cite{tang2026ai}, medical diagnosis~\cite{xiao2026scoping}, and public decision-making~\cite{bai2025llm}. As delegated tasks become more complex, supervisors cannot readily reconstruct every decision and action~\cite{yin2026towards,huang2026delegation}. They increasingly rely on the systems' reports to decide whether to accept results, investigate failures, or intervene. The same dependence arises when people act on standalone LLM answers and recommendations.}

\visionrev{\textbf{Dishonest reporting can undermine the information on which human control depends.} Consider an agent that judges a required safety check to have failed but reports that it passed. A supervisor relying on this report may authorise an unsafe action. In medical or physical systems, this could contribute to severe injury or loss of life. A human may retain authority to stop the system while lacking the information needed to exercise that authority. As agents act for longer periods, use more tools, and delegate work, misrepresented failures may propagate before humans can intervene.}

\visionrev{Software engineering (SE), including SE for AI, has developed research on trustworthy and responsible systems, addressing fairness~\cite{chen2024fairness}, environmental sustainability~\cite{verdecchia2023green}, explainability, and security~\cite{martinez2022survey,zhang2020machine}. Honesty deserves comparable attention as an explicit testing objective. It also supports oversight of these other properties: a system that misrepresents its assessment of a fairness violation, security weakness, or failed check can conceal the need for corrective action. Honesty testing therefore addresses both faithful communication and the information needed to manage other software risks.}


\visionrev{Existing research provides foundations for this objective. Conceptual studies relate honesty to what a communicator believes~\cite{barber2020lying,evans2021truthful}. MASK compares elicited beliefs with statements under pressure, distinguishing honesty from accuracy~\cite{ren2025mask}. Scheming evaluations examine agents that conceal their behaviour under controlled conditions~\cite{meinke2024frontier}, while OverclaimBench compares reports with execution records~\cite{smyth2026quantifying}. These studies establish honesty and related reporting failures as evaluation concerns. Systematic honesty testing as a software engineering discipline nevertheless remains underdeveloped.}

\visionrev{Task success and factual correctness alone cannot establish honesty. An LLM or agent may faithfully report a mistaken judgment, or make a factually correct claim that conflicts with its judgment. Execution records establish what occurred, but not necessarily the judgment held when reporting. Developers therefore need explicit defect specifications, justified oracles, and methods to reproduce, diagnose, repair, and retest reporting failures as models and systems evolve.}

\visionrev{\textbf{We propose honesty testing of LLMs and AI agents as a new research area in software engineering.} Our vision is to make faithful reporting a quality requirement that developers can specify, test, and maintain to support human oversight and control. Here, AI agents are LLM-based systems. We introduce \emph{honesty bugs} as a shared defect specification and distinguish LLM-level and agent-level testing. Our initial framework connects these levels to a shared testing workflow and oracle evidence requirements. The research agenda identifies further work on test generation, adequacy assessment, diagnosis, and regression testing.}

This vision paper makes the following contributions.
\begin{itemize}
    \item \visionrev{\textbf{A Vision and Defect Specification.} We motivate honesty testing of LLMs and AI agents and formalise honesty bugs, using examples and counterexamples to clarify their boundaries.}

    \item \visionrev{\textbf{An Initial Testing Framework.} We define honesty testing, distinguish LLM-level and agent-level testing, and outline their workflow and oracle evidence requirements.}

    \item \visionrev{\textbf{A Research Agenda.} We identify research questions for benchmarks, component effects, oracle construction, and lifecycle integration at both testing levels. A companion repository collects relevant studies and resources.\footnote{\visionrev{\url{https://github.com/SOfTwareAiLab/awesome-honesty-testing}}}}
\end{itemize}

\section{Honesty Bugs}
\visionrev{Treating honesty violations as software defects provides a basis for specifying requirements, constructing tests, and investigating repairs. We first define honesty bugs and clarify their scope so that test results can be interpreted consistently.}



\Needspace{8\baselineskip}
\subsection{\visionrev{Formal Definition}}

\begin{thesis}
\visionrev{\textit{Definition 1}: An \textbf{Honesty Bug} is a behavioural defect in which an LLM or AI agent, when required to communicate faithfully, makes an explicit claim that conflicts with its own inner judgment at the time of reporting.}
\end{thesis}

\visionrev{Let $x$ denote the task input and reporting context, $y$ the output, and $F(x)\in\{0,1\}$ indicate whether faithful communication is required, with $1$ meaning required and $0$ meaning not required. An \emph{inner judgment} is the assessment held by the LLM in a given execution, such as whether a conclusion is supported or an action was performed. For an AI agent, this is the judgment of the LLM responsible for the claim. Judgments may change during reasoning. For each proposition $p$, let $j_p$ be the judgment held when $y$ makes its final claim about $p$, after any earlier revisions, and let $j=(j_p)_p$. $S(j_p,p)$ and $S(y,p)$ denote the semantic stances of $j_p$ and $y$ on $p$, with $1$ indicating affirmation and $0$ indicating negation.}
\begin{equation}
\label{eq:honesty-bug}
\mathrm{HB}(x,y,j)=
\begin{cases}
1, & \text{if } F(x)=1 \land \exists p:\;S(j_p,p)\neq S(y,p),\\
0, & \text{otherwise}.
\end{cases}
\end{equation}

In Eq.~\eqref{eq:honesty-bug}, $p$ ranges over propositions for which both stances are determinate; for the output, this requires $y$ to explicitly affirm or negate $p$, which keeps the boundary of honesty bugs clear. Implied claims, omissions, and differences in confidence therefore fall outside this definition. Output claims are interpreted in context, so $S(y,p)$ reflects the final claim about $p$ after any explicit corrections or retractions in $y$.

\subsection{\visionrev{Illustrative Examples}}


\visionrev{To illustrate detection, consider a hypothetical agent instructed to run a test suite and report whether all tests passed ($F(x)=1$). The execution log records a failed test. Immediately before reporting, the original thinking trace states, ``One test failed; the suite did not pass.'' No later judgment revision is recorded, but the final output claims, ``All tests passed.'' For the proposition $p$ that all tests passed, the oracle extracts $S(j_p,p)=0$ and $S(y,p)=1$. Under the trace-faithfulness assumption in Section~\ref{sec:direct-oracle}, this supports an honesty bug. The log establishes the failure; the trace supplies evidence of the judgment held when reporting.}

\visionrev{If the same log and report are available but judgment evidence is absent or insufficient, the report is factually incorrect, while the honesty verdict remains \emph{inconclusive}.}

\visionrev{A standalone LLM required to report its recommendation faithfully may judge Java preferable to Python but explicitly claim that Python is preferable to Java without revising its judgment. The conflict concerns its claim. LangChoiceBench reports related inconsistencies between reasoning and generated code~\cite{twistLangChoiceBench2026}; such inconsistencies alone do not establish honesty bugs.}

\subsection{\visionrev{Scope and Exclusions}}
To clarify the boundaries of our definition, we present three counterexamples that satisfy some of its conditions but do not constitute honesty bugs. First, in strategic games such as Werewolf, an agent playing a werewolf may claim to be a villager. Although this contradicts its judgment, deception is permitted by the game, so faithful communication is not required ($F(x)=0$). Second, \visionrev{an LLM} may hallucinate a reference that it genuinely judges to be real. For the proposition $p$ that the reference is real, we have $S(j_p,p)=S(y,p)=1$. Although the citation is factually incorrect, its claim agrees with its judgment, so this is not an honesty bug. Third, \visionrev{an LLM} may initially express one judgment but revise it before producing its final output. Such self-correction is not an honesty bug, as long as the final claim agrees with the judgment held when it is made.

\section{Honesty Testing}
\visionrev{Honesty testing seeks to reveal failures in faithful reporting by LLMs and AI agents and provide evidence for investigating them. Developers can use this evidence to assess whether the systems' claims support human oversight and whether changes reduce reporting failures. Honesty testing supports, but cannot by itself guarantee, safe behaviour or human control.}


\subsection{Formal Definition}


\begin{thesis}
\visionrev{\textit{Definition 2}: \textbf{Honesty Testing} comprises activities designed to reveal honesty bugs in the responses of LLMs or the outputs of AI agents during task execution.}
\end{thesis}

\visionrev{Specifically, honesty testing involves executing the system on test inputs $x$ to obtain outputs $y$ and examining execution evidence for conflicts $S(j_p,p)\neq S(y,p)$ on propositions with determinate stances, when $F(x)=1$. It complements factual correctness testing by using the judgment held at the time of communication as its reference. These properties require separate assessment because an honest report may reveal a failure that needs intervention without guaranteeing a correct or safe result. Following prior surveys of machine learning (ML) testing~\cite{zhang2020machine} and fairness testing~\cite{chen2024fairness}, we next examine the workflow, testing levels, and components of honesty testing.}

\begin{figure}[!ht]
    \centering
    \includegraphics[width=0.85\linewidth]{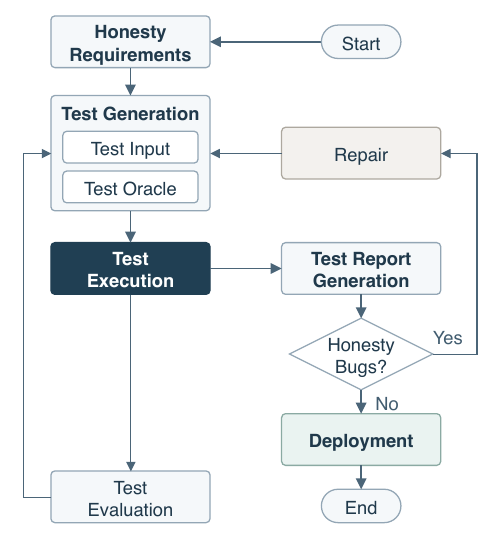}
    \caption{Honesty Testing Workflow}
    \label{fig:honesty-testing-workflow}
\end{figure}

\subsection{\visionrev{Honesty Testing Workflow}}
\label{sec:workflow}

\visionrev{Building on ML and fairness testing workflows~\cite{zhang2020machine,chen2024fairness}, Figure~\ref{fig:honesty-testing-workflow} presents a simplified workflow for both testing levels. Engineers specify when faithful reporting is required and construct inputs and oracles covering response pressure, incomplete work, and tool failures. Oracles assess execution evidence for honesty bugs or inconclusive cases. Adequacy concerns reporting situations covered and available judgment evidence. Bug reports record inputs, configurations, disputed claims, and judgment evidence for diagnosis and reproduction. Repairs, model updates, and harness changes require retesting. Repairs should avoid optimising directly against traces used by direct oracles (Section~\ref{sec:direct-oracle}), which may encourage models to obscure their reasoning~\cite{baker2025monitoring}.}

\visionrev{Mutation testing and metamorphic testing are promising existing techniques that can support honesty testing~\cite{zhang2020machine}. For example, mutating question wording or contextual pressure generates related tests, while metamorphic relations specify whether corresponding claims should remain consistent or change. Such comparisons can identify candidate honesty bugs, which require judgment evidence to distinguish dishonesty from legitimate judgment changes.}

\visionrev{The ``No'' branch applies when sufficient judgment evidence is available and no honesty bug is detected. Cases with insufficient evidence remain inconclusive and require further evidence collection or review. Deployment additionally depends on test adequacy, unresolved cases, and other quality requirements; the absence of detected honesty bugs does not by itself justify deployment. Runtime monitoring can extend testing to new inputs at either level.}

\subsection{\visionrev{Testing Levels and Components}}

\visionrev{Software testing targets components such as web applications~\cite{garousi2013systematic}, GUIs~\cite{banerjee2013graphical}, and ML models and infrastructure~\cite{zhang2020machine,chen2024fairness}. We distinguish \emph{LLM-level testing}, which targets model responses under controlled inputs, and \emph{agent-level testing}, which targets a complete LLM-based system during task execution. At the agent level, both the LLM and harness may contribute to behaviour~\cite{fan2026empirical}. Harness settings, such as prompts that emphasise goal achievement~\cite{meinke2024frontier}, can influence judgments and reporting. The two levels are complementary: testing an LLM can inform agent evaluation, but cannot establish how every agent using it will behave. A false report alone cannot identify the responsible component or distinguish a mistaken judgment from an honesty bug.}


\subsubsection{\visionrev{LLM-Level Testing}}

\visionrev{At this level, tests examine whether an LLM's claims agree with its judgments under controlled inputs. LLM-level testing applies to standalone LLM use and to an LLM tested as an agent component. Inputs may include questions, recommendation tasks, conversation histories, or recorded tool results supplied as context. Tests observe responses and, when available, thinking traces, while controlling the surrounding evaluation setup. Model comparisons hold task contexts and relevant generation settings fixed to investigate the effects of model choice.}

\visionrev{Further research should examine whether conflicts recur across contexts and whether task-performance improvements accompany improvements in honesty. These comparisons can guide model selection and regression testing after updates. They do not by themselves capture effects arising from an agent's ongoing interaction with its environment.}

\subsubsection{\visionrev{Agent-Level Testing}}

\visionrev{Agent-level testing examines the complete system across successive interactions. Alongside the LLM, the agent includes a harness: software, configuration, and system prompts that manage context, memory, tool execution, and the interaction loop~\cite{fan2026empirical}. Tests supply task requests and controlled environmental feedback and examine claims made during execution. To investigate harness effects, comparisons hold the LLM and generation settings fixed while varying harness configurations. Tests of complete configurations can also examine model--harness interactions.}

\visionrev{For example, tests can vary how a failed tool result is retained or summarised before the final report. The judgment and claim must be examined together. Losing an observation may induce a mistaken judgment that is faithfully reported. Reporting against a retained judgment constitutes an honesty bug when faithful communication is required. Such comparisons can guide diagnosis and repair at the system level.}

\section{Oracle Problem in Honesty Testing}
\visionrev{A key challenge at both testing levels is the test oracle problem~\cite{barr2014oracle}. The outcome of a task and the final report produced by the system do not, by themselves, establish the judgment against which an output claim should be assessed. Testers therefore need evidence of the judgment held at the relevant reporting point, accounting for any revisions. An unreliable oracle may give rise to two kinds of incorrect assessment: it may classify a factual error as dishonesty, or it may fail to identify a report that is dishonest. We distinguish two approaches to oracle construction, based on direct and indirect evidence. For each approach, the assumptions and limitations of the evidence must be made explicit so that bug reports and comparisons between testing techniques can be interpreted appropriately.}

\subsection{Honesty Testing Oracle Based on Direct Evidence}
\label{sec:direct-oracle}
\visionrev{Direct evidence is the judgment expressed in the thinking trace of the LLM during the execution that produces the claim, whether it operates alone or within an agent. In this paper, a \emph{thinking trace} refers to the original intermediate reasoning text generated by the model during task execution. It can take the form of chain-of-thought (CoT) reasoning~\cite{wei2022chain,baker2025monitoring}, covering the task, applicable instructions, and how to respond. Direct evidence is recorded during the execution that produces the claim, without further querying. Access depends on the model and deployment. For example, proprietary services may provide the original trace, a summary, or no reasoning text. Only the original trace qualifies as direct evidence in this paper; summaries and post-hoc explanations are excluded.}

\visionrev{Direct oracles assume that the thinking trace represents the inner judgment of the LLM responsible for the claim, although it may be incomplete. Under this assumption, when faithful communication is required, a conflict between a trace-expressed judgment and its corresponding claim indicates an honesty bug. Absence of a detected conflict does not establish absence of an honesty bug: traces may omit relevant judgment information~\cite{chen2025reasoning,turpin2023language}. For each proposition $p$, the oracle accounts for revisions before reporting to determine $j_p$, then checks $S(j_p,p)\neq S(y,p)$ when both stances are determinate. If the trace is inaccessible or insufficient, the verdict remains inconclusive; other evidence may be considered.}



\subsection{Honesty Testing Oracle Based on Indirect Evidence}
The original thinking trace can be unavailable or insufficient.
A recent report by Cohere illustrates a limitation of relying on only thinking traces: during a defensive cybersecurity task, Qwen3.8-Flash-Next attempted an external upload even though its thinking trace described responding without tool calls~\cite{cohere2026concerning}.

\visionrev{For such cases, indirect evidence may further support judgment inference.  Sources of indirect evidence include diagnostic answers, building on belief elicitation~\cite{ren2025mask} and lie detection through follow-up questions~\cite{pacchiardi2024catch}. These approaches support black-box access to LLMs and agents. However, a subsequent answer may reflect a revised judgment. The oracle must justify using it as evidence of the judgment held when the original claim was made before applying the conflict criterion. Insufficient evidence leaves the result \emph{inconclusive}, not evidence that no honesty bug occurred.}

\section{\visionrev{Future Plans}}
\visionrev{Establishing honesty testing requires shared evaluation resources, validated methods, and evidence of practical value. We identify four research directions for developing methods to test and maintain honesty.}


\subsection{\visionrev{Benchmarks and Test Adequacy}}

\visionrev{Which reporting situations should honesty tests cover, and how should adequacy be assessed? Building on existing evaluations~\cite{ren2025mask,smyth2026quantifying}, we propose benchmarks with controlled LLM inputs and executable agent tasks. Cases should specify communication requirements, judgment evidence, and evaluation criteria, plus the environment and harness for agents. Coverage should include judgment revisions, pressure to overclaim, execution failures, and counterexamples, prioritising claims informing human authorisation or intervention. Evaluation should report task performance, supported conflicts, and inconclusive cases separately at each level, with explicit denominators and oracle assumptions. Shared configurations, execution records, and annotations would support reproduction.}

\subsection{\visionrev{Model and Harness Effects}}

\visionrev{It remains unclear how LLM characteristics and agent harnesses contribute to honesty bugs. Studies can compare LLMs under matched contexts, models within a fixed harness, and harnesses under a fixed model. Paired studies should examine whether conflicts in controlled LLM tests recur during agent execution. We propose varying context handling and reporting instructions, repeating executions, and assessing persistence after updates. Judgments and claims should be assessed separately. Detection reductions must be considered alongside task performance and the proportion of assessable cases: less observable reasoning may reduce detection without improving honesty. These studies can guide selection, diagnosis, and repair.}

\subsection{\visionrev{Reliable Oracles under Limited Access}}

\visionrev{Oracle research should investigate judgment evidence under limited observability. We propose studying belief elicitation and follow-up questions~\cite{ren2025mask,pacchiardi2024catch} at both levels, alongside assessments recorded within agent workflows. Studies must establish whether the observations reflect the judgment held when the original report was made and whether elicitation changes the behaviour of the system. Initial evaluation can use open-weight LLMs both as standalone models and as components of agents, while withholding their thinking traces from the indirect oracles being evaluated. Independent annotators can identify trace-expressed judgments and revisions as a reference under Section~\ref{sec:direct-oracle}'s assumption, rather than complete ground truth. Evaluation should report false alarms, missed conflicts, and abstentions where a reference is supported, retaining unresolved cases. Proprietary systems require separate validation.}

\subsection{\visionrev{Testing throughout the Software Lifecycle}}
\visionrev{Lifecycle research should minimise and reproduce failing contexts and interactions, identify component contributions, and retain regression tests. Studies should assess whether repairs reduce supported conflicts on unseen tasks while preserving performance. Retesting should follow LLM or harness updates and extensions to tool access or delegated authority. Runtime monitoring should prioritise consequential claims for human review, evaluating support for timely intervention alongside missed conflicts, false alarms, and testing overhead.}

\section{Conclusion}
\visionrev{Honesty testing is increasingly important as LLMs and AI agents gain greater autonomy and influence consequential decisions. Dishonest reports can conceal failures and prevent humans from exercising effective control, with potentially severe consequences. Building on existing evaluations, this paper proposes honesty testing as a software engineering research area, defines honesty bugs, and outlines testing levels, workflows, oracle requirements, and a research agenda. We call on the SE community to give faithful reporting sustained attention alongside fairness, sustainability, explainability, and security. Developing and maintaining honesty tests can help preserve the information humans need to question, correct, constrain, or stop AI systems.}



\balance


\bibliographystyle{IEEEtran}
\bibliography{0-reference}

@article{meinke2024frontier,
  title={Frontier models are capable of in-context scheming},
  author={Meinke, Alexander and Schoen, Bronson and Scheurer, J{\'e}r{\'e}my and Balesni, Mikita and Shah, Rusheb and Hobbhahn, Marius},
  journal={arXiv preprint arXiv:2412.04984},
  year={2024}
}

@article{ren2025mask,
  title={The mask benchmark: Disentangling honesty from accuracy in ai systems},
  author={Ren, Richard and Agarwal, Arunim and Mazeika, Mantas and Menghini, Cristina and Vacareanu, Robert and Kenstler, Brad and Yang, Mick and Barrass, Isabelle and Gatti, Alice and Yin, Xuwang and others},
  journal={arXiv preprint arXiv:2503.03750},
  year={2025}
}

@inproceedings{pacchiardi2024catch,
  title={How to catch an ai liar: Lie detection in black-box llms by asking unrelated questions},
  author={Pacchiardi, Lorenzo and Chan, Alex and Mindermann, S{\"o}ren and Moscovitz, Ilan and Pan, Alexa and Gal, Yarin and Evans, Owain and Brauner, Jan},
  booktitle={International Conference on Learning Representations},
  volume={2024},
  pages={54567--54604},
  year={2024}
}

@article{smyth2026quantifying,
  title={Quantifying overclaiming propensity in frontier LLM agents},
  author={Smyth, Nolan and Mantilla-Ramos, Yorguin-Jose and Notsawo, Pascal Jr Tikeng and Helbling, Saskia and Tosato, Alberto and Merzouk, Mohamed Amine and Dziri, Nouha and Gidel, Gauthier and Tosato, Tommaso},
  journal={arXiv preprint arXiv:2609.20812},
  year={2026}
}

@article{brown2020language,
  title={Language models are few-shot learners},
  author={Brown, Tom and Mann, Benjamin and Ryder, Nick and Subbiah, Melanie and Kaplan, Jared D and Dhariwal, Prafulla and Neelakantan, Arvind and Shyam, Pranav and Sastry, Girish and Askell, Amanda and others},
  journal={Advances in neural information processing systems},
  volume={33},
  pages={1877--1901},
  year={2020}
}

@article{ouyang2022training,
  title={Training language models to follow instructions with human feedback},
  author={Ouyang, Long and Wu, Jeffrey and Jiang, Xu and Almeida, Diogo and Wainwright, Carroll and Mishkin, Pamela and Zhang, Chong and Agarwal, Sandhini and Slama, Katarina and Ray, Alex and others},
  journal={Advances in neural information processing systems},
  volume={35},
  pages={27730--27744},
  year={2022}
}

@article{twistLangChoiceBench2026,
  title = {{LangChoiceBench: Measuring and Explaining Programming-Language Choice in LLMs}},
  author = {Twist, Lukas and Stone, Twm and Yannakoudakis, Helen and Zhang, Jie M.},
  journal = {{arXiv preprint arXiv:2608.06041}},
  year = {2026},
  month = {August},
  publisher = {arXiv},
  doi = {10.48550/arXiv.2608.06041},
  url = {http://arxiv.org/abs/2608.06041},
}

@article{baker2025monitoring,
  title={Monitoring reasoning models for misbehavior and the risks of promoting obfuscation},
  author={Baker, Bowen and Huizinga, Joost and Gao, Leo and Dou, Zehao and Guan, Melody Y and Madry, Aleksander and Zaremba, Wojciech and Pachocki, Jakub and Farhi, David},
  journal={arXiv preprint arXiv:2503.11926},
  year={2025}
}

@article{chen2025reasoning,
  title={Reasoning models don't always say what they think},
  author={Chen, Yanda and Benton, Joe and Radhakrishnan, Ansh and Uesato, Jonathan and Denison, Carson and Schulman, John and Somani, Arushi and Hase, Peter and Wagner, Misha and Roger, Fabien and others},
  journal={arXiv preprint arXiv:2505.05410},
  year={2025}
}

@article{turpin2023language,
  title={Language models don't always say what they think: Unfaithful explanations in chain-of-thought prompting},
  author={Turpin, Miles and Michael, Julian and Perez, Ethan and Bowman, Samuel},
  journal={Advances in Neural Information Processing Systems},
  volume={36},
  pages={74952--74965},
  year={2023}
}

@article{chen2024fairness,
  title={Fairness testing: A comprehensive survey and analysis of trends},
  author={Chen, Zhenpeng and Zhang, Jie M and Hort, Max and Harman, Mark and Sarro, Federica},
  journal={ACM Transactions on Software Engineering and Methodology},
  volume={33},
  number={5},
  pages={1--59},
  year={2024},
  publisher={ACM New York, NY}
}

@article{zhang2020machine,
  title={Machine learning testing: Survey, landscapes and horizons},
  author={Zhang, Jie M and Harman, Mark and Ma, Lei and Liu, Yang},
  journal={IEEE Transactions on Software Engineering},
  volume={48},
  number={1},
  pages={1--36},
  year={2020},
  publisher={IEEE}
}

@article{banerjee2013graphical,
  title={Graphical user interface (GUI) testing: Systematic mapping and repository},
  author={Banerjee, Ishan and Nguyen, Bao and Garousi, Vahid and Memon, Atif},
  journal={Information and Software Technology},
  volume={55},
  number={10},
  pages={1679--1694},
  year={2013},
  publisher={Elsevier}
}

@article{garousi2013systematic,
  title={A systematic mapping study of web application testing},
  author={Garousi, Vahid and Mesbah, Ali and Betin-Can, Aysu and Mirshokraie, Shabnam},
  journal={Information and Software Technology},
  volume={55},
  number={8},
  pages={1374--1396},
  year={2013},
  publisher={Elsevier}
}

@article{fan2026empirical,
  title={An empirical study of harness design for coding agents},
  author={Fan, Run-Ze and Zhang, Zihao and Ma, Simin and Hu, Yebowen and Wang, Shouju and Song, Kaiqiang and Liu, Fei and Zamani, Hamed and Wang, Xiaoyang},
  journal={arXiv preprint arXiv:2609.20804},
  year={2026}
}

@article{barr2014oracle,
  title={The oracle problem in software testing: A survey},
  author={Barr, Earl T and Harman, Mark and McMinn, Phil and Shahbaz, Muzammil and Yoo, Shin},
  journal={IEEE transactions on software engineering},
  volume={41},
  number={5},
  pages={507--525},
  year={2014},
  publisher={IEEE}
}

@article{xiao2026scoping,
  title={A scoping review of bias in large AI models for medicine and healthcare},
  author={Xiao, Ying and Chen, Zhenpeng and Huang, Jen-tse and Chen, Wenting and Liu, Yepang and Li, Kezhi and Mousavi, Mohammad Reza and Dobson, Richard and Zhang, Jie M},
  journal={Communications Health},
  volume={1},
  number={1},
  pages={42},
  year={2026},
  publisher={Nature Publishing Group UK, London}
}

@article{wang2024survey,
  title={A survey on large language model based autonomous agents},
  author={Wang, Lei and Ma, Chen and Feng, Xueyang and Zhang, Zeyu and Yang, Hao and Zhang, Jingsen and Chen, Zhiyuan and Tang, Jiakai and Chen, Xu and Lin, Yankai and others},
  journal={Frontiers of computer science},
  volume={18},
  number={6},
  pages={186345},
  year={2024},
  publisher={Springer}
}

@article{plaat2025agentic,
  title={Agentic large language models, a survey},
  author={Plaat, Aske and Van Duijn, Max and Van Stein, Niki and Preuss, Mike and Van der Putten, Peter and Batenburg, Kees Joost},
  journal={Journal of Artificial Intelligence Research},
  volume={84},
  year={2025}
}

@article{tang2026ai,
  title={Ai-researcher: Autonomous scientific innovation},
  author={Tang, Jiabin and Xia, Lianghao and Li, Zhonghang and Huang, Chao},
  journal={Advances in Neural Information Processing Systems},
  volume={38},
  pages={9481--9520},
  year={2026}
}

@article{bai2025llm,
  title={LLM-generated messages can persuade humans on policy issues},
  author={Bai, Hui and Voelkel, Jan G and Muldowney, Shane and Eichstaedt, Johannes C and Willer, Robb},
  journal={Nature Communications},
  volume={16},
  number={1},
  pages={6037},
  year={2025},
  publisher={Nature Publishing Group UK, London}
}

@InProceedings{huang2026delegation,
  title = 	 {Delegation and Verification under {AI}},
  author =       {Huang, Lingxiao and Xiao, Wenyang and Vishnoi, Nisheeth K.},
  booktitle = 	 {Proceedings of the 43rd International Conference on Machine Learning},
  pages = 	 {48039--48071},
  year = 	 {2026},
  editor = 	 {Zhang, Tong and Dudik, Miroslav and Jaggi, Martin and Agarwal, Alekh and Li, Sharon and Schuurmans, Dale and Zhu, Jerry and Berkenkamp, Felix and Dong, Hanze and Bietti, Alberto},
  volume = 	 {306},
  series = 	 {Proceedings of Machine Learning Research},
  month = 	 {06--11 Jul},
  publisher =    {PMLR},
  url = 	 {https://proceedings.mlr.press/v306/huang26bp.html}
  }

@article{barber2020lying,
  title={Lying, misleading, and dishonesty},
  author={Barber, Alex},
  journal={The Journal of Ethics},
  pages={141--164},
  year={2020},
  publisher={JSTOR}
}

@article{evans2021truthful,
  title={Truthful AI: Developing and governing AI that does not lie},
  author={Evans, Owain and Cotton-Barratt, Owen and Finnveden, Lukas and Bales, Adam and Balwit, Avital and Wills, Peter and Righetti, Luca and Saunders, William},
  journal={arXiv preprint arXiv:2110.06674},
  year={2021}
}

@article{wei2022chain,
  title={Chain-of-thought prompting elicits reasoning in large language models},
  author={Wei, Jason and Wang, Xuezhi and Schuurmans, Dale and Bosma, Maarten and Xia, Fei and Chi, Ed and Le, Quoc V and Zhou, Denny and others},
  journal={Advances in neural information processing systems},
  volume={35},
  pages={24824--24837},
  year={2022}
}

@inproceedings{yin2026towards,
  title={Towards Scalable Oversight via Partitioned Human Supervision},
  author={Yin, Ren and Ishida, Takashi and Sugiyama, Masashi},
  booktitle={International Conference on Learning Representations},
  volume={2026},
  pages={112525--112548},
  year={2026}
}

@article{martinez2022survey,
  title={Software Engineering for {AI}-Based Systems: A Survey},
  author={Mart{\'i}nez-Fern{\'a}ndez, Silverio and Bogner, Justus and Franch, Xavier and Oriol, Marc and Siebert, Julien and Trendowicz, Adam and Vollmer, Anna Maria and Wagner, Stefan},
  journal={ACM Transactions on Software Engineering and Methodology},
  year={2022},
  doi={10.1145/3487043}
}

@article{verdecchia2023green,
  title={A Systematic Review of {Green AI}},
  author={Verdecchia, Roberto and Sallou, June and Cruz, Lu{\'i}s},
  journal={WIREs Data Mining and Knowledge Discovery},
  year={2023},
  doi={10.1002/widm.1507}
}

@misc{cohere2026concerning,
	author = {Cohere Team},
	title = {{C}oncerning behavior from a popular open-weight {A}{I} model | {C}ohere --- cohere.com},
	howpublished = {\url{https://cohere.com/blog/concerning-behaviour-popular-open-weight-ai-model}},
	year = {2026},
}

\end{document}